\documentclass[prl,twocolumn,showpacs,preprintnumbers,amsmath,amssymb]{revtex4}
\usepackage{graphicx}
\usepackage{dcolumn}
\usepackage{bm}
\usepackage{psfig}
\newcommand {\be}{\begin{eqnarray}}
\newcommand {\ee}{\end{eqnarray}}

\begin{document}
\title{Phase diagram and binding energy of interacting Bose gases}
\author{M. M{\"a}nnel$^{1}$, K. Morawetz$^{2,3}$, M. Schreiber$^{1}$ and P. Lipavsk\'y$^{4,5}$}
\affiliation{$^1$Institute of Physics, Chemnitz University of Technology, 09107 Chemnitz, Germany}
\affiliation{$^2$Research-Center Dresden-Rossendorf, Bautzner Landstr. 128, 01328 Dresden, Germany}
\affiliation{$^3$Max-Planck-Institute for the Physics of Complex Systems, N{\"o}thnitzer Str. 38, 01187 Dresden, Germany}
\affiliation{$^4$Institute of Physics, Academy of Sciences, Cukrovarnick\'a 10, 16253 Prague 6, Czech Republic}
\affiliation{$^5$Faculty of Mathematics and Physics, Charles University, Ke Karlovu 3, 12116 Prague 2, Czech Republic}

\begin{abstract}
From the many-body T-matrix the condition for a medium-dependent bound state and its binding energy is derived for a homogeneous interacting Bose gas. This condition provides the critical line in the phase diagram in terms of the medium-dependent scattering length. Separating the Bose pole from the distribution function the influence of a Bose condensate is discussed and a thermal minimum of the critical scattering length is found.
\end{abstract}
\date{\today}
\pacs{05.30.Jp, 03.75.Hh, 05.30.-d, 64.60.Fr}
\maketitle

\begin{figure}[t]
\psfig{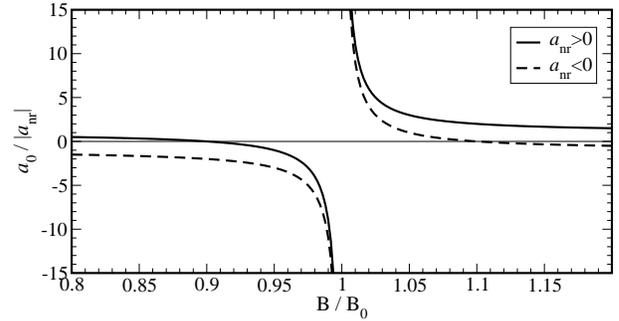} 
\caption{Scattering length in the vicinity of a Feshbach resonance}
\label{ares}
\end{figure}

The discovery of Feshbach resonances in gases of ultra-cold bosons in 1998 \mbox{\cite{IASMSK98,CFHAV98,RCBGCW98}} has provided an important tool to analyze interacting Bose systems. Near these resonances it is possible to tune the interaction and especially the free scattering length $a_0$ with an external magnetic field $B$. In the vicinity of the resonance at $B=B_0$ the scattering length is \cite{PS04}
\be
a_0=a_{\rm nr}\left(1+\frac{\Delta B}{B-B_0}\right),
\ee
as shown in Fig.~\ref{ares}. $a_{\rm nr}$ is the scattering length far away from the resonance and $\Delta B\propto1/a_{\rm nr}$ describes the width of the resonance. A Fermi gas near a Feshbach resonance can be driven through a transition from a Bose-Einstein condensate (BEC) of two-particle bound states for $a_0>0$ to a BCS state of Cooper pairs for $a_0<0$ \cite{RGJ04}. For bosons the influence of the interaction on the Bose condensation is of main interest. Interacting Bose gases at ultra-low temperatures are expected to consist of unbound, bound and condensed bosons, furthermore one expects an influence of the interaction on the critical temperature and density of Bose condensation, see citations in \cite{MMS07}. Here we focus on the formation of bound states in the presence or absence of a Bose condensate. We will derive the condition for bound states in terms of the medium-dependent scattering length to discuss the phase diagram and the binding energy.

The two-particle scattering is described with the many-body T-matrix in ladder and quasi-particle approximation
\begin{align}
{\cal T}_{p\bar p}\!\left(\!Q,\omega\!\right)\!=\! {\cal V}_{p\bar p}
\!+\!\!\! \int \!\!\!\frac{d^3q}{\left(2\pi\right)^3} {\cal T}_{pq}\!\left(\!Q,\omega\!\right)\! \frac{1\!+\!f_{\frac{Q}{2}-q}\!+\!f_{\frac{Q}{2}+q}}{\omega\!-\!\frac{\hbar^2 Q^2}{4m}\!-\!\frac{\hbar^2q^2}{m}\!+\!i\eta}  {\cal V}_{q\bar p}.
\label{bse}
\end{align}
The influence of the surrounding particles is represented by the distribution function $f_p$. The total momentum $Q$ reflects the center-of-mass motion of the scattering particles relative to the medium, while $q$ is their relative momentum. The medium is assumed to be a homogeneous ideal Bose gas with a distribution \cite{ST97}
\be
f_p=\frac{1}{e^{\left(\frac{\hbar^2p^2}{2m}-\mu\right)/T}-1}+\frac{(2\pi)^3 n_0}{2F+1} \delta(p)
\label{f}
\ee
and a density
\be
n=(2F+1)\int\frac{d^3p}{(2\pi)^3}\frac{1}{e^{\left(\frac{\hbar^2p^2}{2m}-\mu\right)/T}-1}+n_0,
\label{n}
\ee
where $n_0$ is the condensate density and $F$ is the total spin. Since we assume that the interaction does not change the spin there are no degradation factors in the T-matrix (\ref{bse}). In the normal state $n_0=0$ and Eq.~(\ref{n}) yields the chemical potential $\mu$ as a function of the temperature and density. In the superfluid state $\mu=0$ and Eq.~(\ref{n}) determines the condensate density $n_0$. At the critical point $\mu=0$ and $n_0=0$ and from (\ref{n}) follow the critical temperature $T_{\rm C}$ and the critical density $n_{\rm C}$. The dependence of the critical properties on the interaction will be neglected.

At low temperatures only s-wave scattering at small momenta is important, furthermore we want to concentrate on bound states near the continuum edge. Therefore and for the sake of simplicity we neglect the range of the interaction and assume the interaction to be a contact interaction for which the potential is independent of the relative momenta, ${\cal V}_{pq}={\cal V}$. Accordingly, the \mbox{T-matrix} is also independent of the relative momenta, \mbox{${\cal T}_{pq}(Q,\omega)={\cal T}(Q,\omega)$}, and Eq. (\ref{bse}) simplifies to the algebraic relation \mbox{${\cal T}={\cal V}/(1-{\cal G}{\cal V})$} with the two-particle progator
\be
{\cal G}\left(Q,\omega\right)=\int\frac{d^3q}{\left(2\pi\right)^3} \frac{1+f_{\frac{Q}{2}-q}+f_{\frac{Q}{2}+q}}{\omega-\frac{\hbar^2 Q^2}{4m}-\frac{\hbar^2q^2}{m}+i\eta}.
\label{G}
\ee
Obviously we can split ${\cal G}$ into the free propagator ${\cal G}_0$ resulting in the $f\to0$ limit and the medium correction ${\cal G}_{\rm m}\propto f$. The free propagator diverges and a cutoff is necessary. To circumvent this cutoff, we introduce the vacuum T-matrix ${\cal T}_0={\cal V}/(1-{\cal G}_0{\cal V})$, which determines the free scattering length
\be
a_0=\frac{m}{4\pi\hbar^2}{\cal T}_0\left(Q,\frac{\hbar^2Q^2}{4m}\right).
\label{a0T0}
\ee
The strength of the interaction is now described by the free scattering length, which is interpreted as the relevant physical quantity tunable near the Feshbach resonance.

Solving (\ref{a0T0}) yields
\be
a_0=\frac{m}{4\pi\hbar^2}\lim_{q_0\to\infty}\frac{{\cal V}}{1+\frac{m}{2\pi^2\hbar^2}{\cal V}\int\limits_0^{q_0}dq}.
\label{a0}
\ee
The inverse of the cutoff $q_0$ is proportional to the range of the interaction, i.e., contact interaction means $q_0\to\infty$. Postulating a finite $a_0$, it is necessary to renormalize the interaction strength \cite{MMS07,SMR97,PS00,BPYZ07}, such that
\be
{\cal V}
=\lim_{q_0\to \infty} \left(-\frac{2\pi^2\hbar^2}{mq_0}\frac{1}{1-\frac{\pi}{2a_0q_0}}\right)
\label{g}
\ee
follows from (\ref{a0}). Frequently the pseudopotential \cite{PS04,BW59}
\be
\bar {\cal V}=\frac{4\pi\hbar^2a_0}{m}
\label{pseudo}
\ee
is used to describe the dependence of the interaction strength on the scattering length for contact interaction \cite{DGPS99,BBHLV01}. The difference to (\ref{g}) is that the leading term of (\ref{g}) with respect to a large cutoff $q_0$ is always negative and independent of $a_0$, i.e., the contact interaction is always attractive, while the pseudopotential is attractive for $a_0<0$ and repulsive for $a_0>0$. The reason for this difference is that the pseudopotential is only valid up to first-order Born approximation \cite{BW59}. Within this approximation one neglects ${\cal G}$, i.e., ${\cal T}={\cal V}$, such that the diverging terms in the denominators of (\ref{a0}) and (\ref{g}) vanish.

\begin{figure}[t]
\parbox[t]{9.cm}{
\parbox[b]{4.4cm}{
\flushleft{a)}
\psfig{file=ebscnnk8.eps,width=4.4cm}}
\parbox[b]{4.4cm}{
\flushleft{b)}
\psfig{file=ebsctnk8.eps,width=4.4cm}}
}
\parbox[t]{9.cm}{
\parbox[b]{4.4cm}{
\flushleft{c)}
\psfig{file=ebscq7.eps,width=4.4cm}}
\parbox[b]{4.4cm}{
\flushleft{d)}
\psfig{file=ebscs3.eps,width=4.4cm}}
}
\centerline{\parbox[t]{4.4cm}{
\flushleft{e)}
\psfig{file=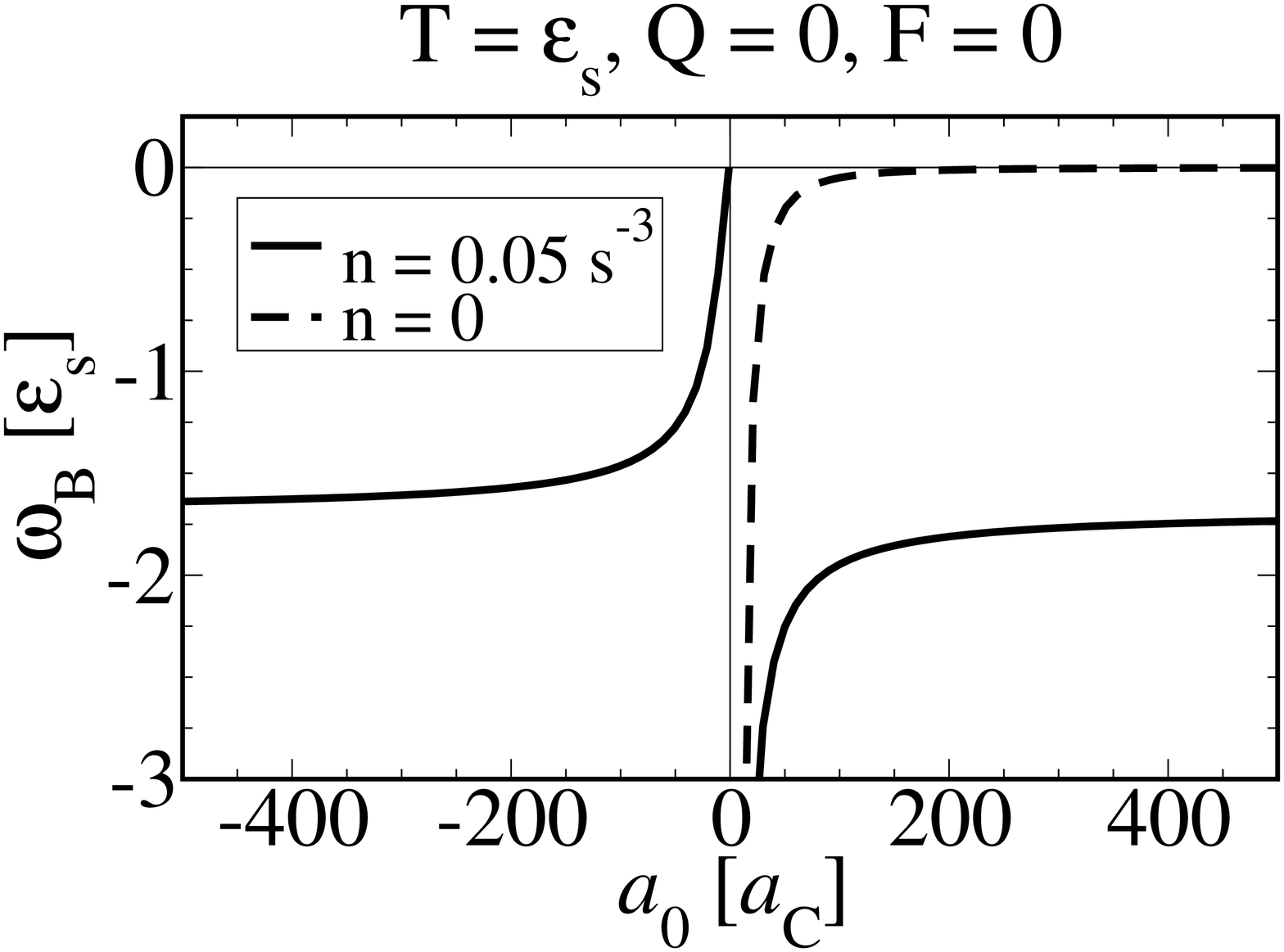,width=4.4cm}
}}
\caption{Binding energy for the bound state of a homogeneous Bose gas with contact interaction. Fixed parameters are given above the plots. \mbox{$n_{\rm C}=0.06s^{-3}$}, \mbox{$T_{\rm C}=0.90\varepsilon_s$}, $a_{\rm C}=0.06s$
\label{ob}}
\end{figure}

The in-medium T-matrix can be expressed by the free one ${\cal T}={\cal T}_0/(1-{\cal G}_{\rm m}{\cal T}_0)$. Accordingly, the in-medium scattering length is
\be
a=\frac{m}{4\pi\hbar^2}{\cal T}\left(Q,\frac{\hbar^2Q^2}{4m}\right)&=&\frac{a_0}{1-\frac{4\pi\hbar^2a_0}{m}{\cal G}_{\rm m}\left(Q,\frac{\hbar^2Q^2}{4m}\right)}\nonumber\\
&=&\frac{a_0}{1+\frac{a_0}{a_{\rm C}}}
\label{a}
\ee
with $a_{\rm C}\ge0$. We use here the definition of the many-body scattering length of \cite{BS96} instead of the definition used in \cite{SMR97,MMS07}. Instead of a divergence the many-body scattering length for $Q=0$ therefore has a zero at the critical point of Bose condensation, i.e., $a_{\rm C}=0$. Splitting (\ref{G}) in another way
\be
{\cal G}\left(Q,\omega\right)={\cal G}\left(Q,\frac{\hbar^2Q^2}{4m}\right)+\frac{m}{4\pi\hbar^2}{\cal J}\left(Q,\omega\right),
\ee
with
\be
{\cal J}\left(Q,\omega+\frac{\hbar^2Q^2}{4m}\right)=\int\!\frac{d^3q}{\left(2\pi\right)^3} \frac{4\pi\omega}{q^2} \frac{1\!+\!f_{\frac{Q}{2}-q}\!+\!f_{\frac{Q}{2}+q}}{\omega\!-\!\frac{\hbar^2q^2}{m}\!+\!i\eta}
\label{J}
\ee
one obtains for the in-medium T-matrix
\be
{\cal T}\left(Q,\omega\right)=\frac{4\pi\hbar^2 a}{m}\frac{1}{1-a{\cal J}\left(Q,\omega\right)}.
\label{T}
\ee
\begin{figure}[t]
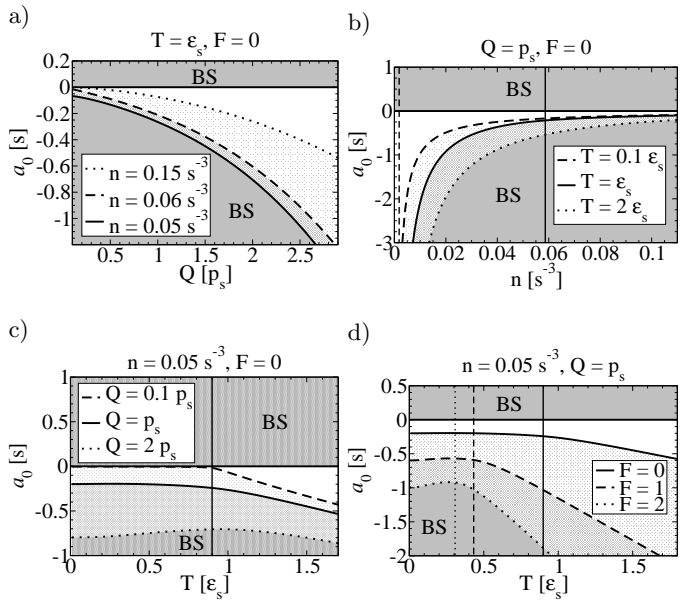

\parbox[t]{9.cm}{
\parbox[b]{4.4cm}{
\flushleft{a)}
\psfig{file=acqqnkengl4.eps,width=4.4cm}}
\parbox[b]{4.4cm}{
\flushleft{b)}
\psfig{file=acnqnkengl10.eps,width=4.4cm}}
}
\parbox[t]{9.cm}{
\parbox[b]{4.4cm}{
\flushleft{c)}
\psfig{file=actqnkengl9.eps,width=4.4cm}}
\parbox[b]{4.4cm}{
\flushleft{d)}
\psfig{file=acs4.eps,width=4.4cm}}
}
\caption{Phase diagram for the appearance of the bound state (BS) of a homogeneous Bose gas with contact interaction. The vertical lines mark the onset of Bose condensation for the corresponding parameters
\label{ac}}
\end{figure}
A bound state, i.e., a pole of the T-matrix, is therefore possible if
\be
0<{\cal J}\left(Q,\omega_{\rm B}+\frac{\hbar^2Q^2}{4m}\right)=\frac{1}{a},
\label{bcond}
\ee
where the corresponding binding energy $\omega_{\rm B}<0$ is measured relative to the continuum edge, i.e., $\frac{\hbar^2Q^2}{4m}$. The binding energy is shown in Fig.~\ref{ob}. The fixed parameters in the plots define the corresponding length scale $s$, i.e., $Q=p_s=1/s$, $T=\varepsilon_s=\hbar^2/2ms^2$ or $n=0.05s^{-3}$, respectively. According to (\ref{a}) the condition for the bound state \mbox{$a>0$} is satisfied in two situations. In the first case $a_0>0$ the interaction is strong enough to form the bound state already in the vacuum. In the second case $a_0<-a_{\rm C}$ the bound state is induced by the medium. For $-a_{\rm C}<a<0$ the interaction is attractive but insufficient to form a bound state. For bosons near a Feshbach resonance this means that in addition to the bound state in the $a_0>0$ region, a bound state is also possible on the other side of the resonance for $a_0<-a_{\rm C}$. The appearance of the medium-induced bound state is also signaled by the divergence of the medium-dependent scattering length (\ref{a}) at $a_0=-a_{\rm C}$ \cite{N02}. From these three cases the dependence of the bound-state region on the density, temperature, total momentum, scattering length and spin follows as shown in Fig.~\ref{ac}.

\begin{figure}[t]
\psfig{file=acinv6.eps,width=8cm}
\psfig{file=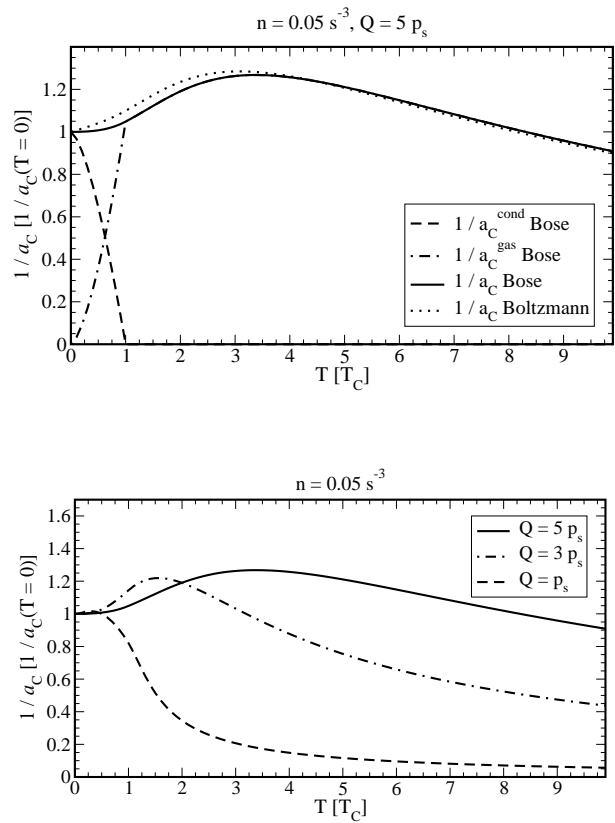,width=8cm}  
\caption{Inverse critical scattering length versus temperature for different approximations (upper plot), and for different total momenta for a Bose gas with condensate (lower plot)}
\label{acinv}
\end{figure}

\begin{table*}[t]
\caption{Binding energies for different species. The singlet (s) and triplet (t) scattering lengths $a_0$ are from \cite{PS04} if not marked differently. The length scale $s$ is chosen so that \mbox{$T\!=\!\varepsilon_s\!=\!0.5\mu{\rm K}\, k_{\rm B}\!\approx\!43{\rm peV}$}. The binding energy $\omega_{\rm B}$ as follows from (\ref{bcond}) for $Q\!=\!0$ and 
\mbox{$n\!=\!0.85 n_{\rm C}$} is compared to its vacuum value $\omega_{\rm B0}=-\mbox{ }\hbar^2/m a_0^2$, $a_{\rm C}\!\approx\!64\!\times\!10^{-3}s$. The total spin $F$ is that of the hyperfine state with lowest energy.
\label{tab}}
\begin{tabular}[t]{|c|c|c|c|c|c|c|c|}
\hline
&&&&&&&\\[-2ex]
& $s\left[{\rm nm}\right]$ & $F$ & $n\left[{\rm cm}^{-3}\right]$ & & $a_0\left[10^{-3} s\right]$ & $\omega_{\rm B}\left[{\rm neV}\right]$ & $\omega_{\rm B0}\left[{\rm neV}\right]$
\cr
\hline
&&&&&&\multicolumn{2}{c|}{}\\[-2ex]
 $^{7}$Li & $263$ & $1$ & $8.2\times10^{12}$ & t & $-5.55\pm0.11$ & \multicolumn{2}{c|}{no bound state}
\cr
\hline
&&&&&&&\\[-2ex]
$^{85}$Rb & $76$  & $2$ & $5.8\times10^{14}$ & t & $-258\pm12$    & $-(52\pm5)\times10^{-4}$ & no bound state
\cr
$^{133}$Cs & $60$ & $3$ & $1.6\times10^{15}$ & t & $ 2100\pm90$   & $-(112.2\pm2.2)\times10^{-3}$ & $-(19.6\pm1.7)\times10^{-3}$
\cr
$^{85}$Rb & $76$  & $2$ & $5.8\times10^{14}$ & s & $ 1700\pm500$  & $-0.133\pm0.023$ & $-0.039\pm0.022$
\cr
$^{133}$Cs & $60$ & $3$ & $1.6\times10^{15}$ & s & $ 245\pm9$     & $-1.49\pm0.11$   & $-1.44\pm0.11$
\cr
$^{87}$Rb & $75$  & $1$ & $3.6\times10^{14}$ & t & $ 75.1\pm2.9$  & $-15.4\pm1.2$    & $-15.3\pm1.2$
\cr
          &       &     &                    & s & $ 63.8\pm0.8$  & $-21.2\pm0.6$    & $-21.2\pm0.6$
\cr
\hline
&&&&&&\multicolumn{2}{c|}{}\\[-2ex]
$^{4}$He  & $348$ & $0$ & $1.2\times10^{12}$ & s & $ 30\pm6$ \cite{GSTHKS00} & \multicolumn{2}{c|}{$-110\pm50$}
\cr
$^{23}$Na & $145$ & $1$ & $4.9\times10^{13}$ & t & $ 23.8\pm0.4$  & \multicolumn{2}{c|}{$-152\pm6$}
\cr
          &       &     &                    & s & $ 7.0\pm0.8$   & \multicolumn{2}{c|}{$-1800\pm500$}
\cr
$^{7}$Li  & $263$ & $1$ & $8.2\times10^{12}$ & s & $ 6.6\pm0.5$   & \multicolumn{2}{c|}{$-2000\pm400$}
\cr
$^{1}$H   & $696$ & $0$ & $1.5\times10^{11}$ & t & $ 9.1\times10^{-2}$ & \multicolumn{2}{c|}{$-1.04\times10^7$}
\cr
          &       &     &                    & s & $ 3.1\times10^{-2}$ & \multicolumn{2}{c|}{$-8.97\times10^7$}
\cr
\hline
\end{tabular}
\end{table*}

For large $|a_0|$ the binding energy converges towards a finite value $\omega_{\rm C}$ as seen in Fig.~\ref{ob}\,e), which follows from \mbox{${\cal J}(Q,\omega_{\rm C}+\hbar^2Q^2/4m)=1/a_{\rm C}$}. This convergence can be explained by the convergence of the medium-dependent scattering length (\ref{a}) and also by the convergence of the interaction strength (\ref{g}) for large $|a_0|$. The interaction strength (\ref{g}) also shows that the interaction is stronger for positive $a_0$ and therefore $|\omega_{\rm B}|$ is larger in this case. The region where a bound state is possible spreads with increasing density, as can be seen in Fig.~\ref{ac}\,a). Fig.~\ref{ob}\,a) shows that for increasing density $|\omega_{\rm B}|$ increases, too. This behavior can be explained by the increasing influence of many-body effects with increasing density. Due to the Bose enhancement the formation of bound states is supported. On the other hand thermal fluctuations hinder the formation of bound states, which is shown by the shrinkage of the bound-state region in Fig.~\ref{ac}\,b) and the decrease of $|\omega_{\rm B}|$ in Fig.~\ref{ob}\,b) with increasing temperature. The motion of the scattered particles relative to the medium has a similar effect as Figs.~\ref{ob}\,c) and \ref{ac}\,c) show. In the limit of vanishing density the bound-state condition is $a_0>0$ and the binding energy is $\omega_{\rm B0}=-\mbox{ }\hbar^2/m a_0^2$. This simple result can only be explained by the potential (\ref{g}). If one would follow  the philosophy of pseudopotentials instead, one would have repulsion for $a_0>0$, and no bound states. In Figs.~\ref{ob}\,d) and \ref{ac}\,d) one sees that with increasing total spin $|\omega_{\rm B}|$ the bound-state region decreases. The reason is that with increasing total spin the density of states also increases and therefore the occupation of states for a given density decreases. The effect is therefore similar to that of a decrease of density.

A remarkable feature in Figs.~\ref{ac}\,c) and d) is that a maximum of the critical line appears at some temperature $T_{\rm ex}$. This means that in these cases the region of bound states becomes smaller if the temperature is decreased further, which is in contrast to the behavior above $T_{\rm ex}$. To understand this effect we observe from (\ref{f}) and (\ref{a}) that in the superfluid state $1/a_{\rm C}$ can be split into two parts, $1/a_{\rm C}=1/a_{\rm C}^{\rm cond}+1/a_{\rm C}^{\rm gas}$. The first part $1/a_{\rm C}^{\rm cond}\propto n_0/Q^2$ bears the contribution from the condensate. The second part $1/a_{\rm C}^{\rm gas}\propto f$ represents the uncondensed Bose gas vanishing as the temperature approaches zero. While $a_{\rm C}^{\rm gas}$ diverges at $T=0$ and $a_{\rm C}^{\rm cond}$ diverges at $T=T_{\rm C}$ the scattering length $a_{\rm C}$ remains finite as Figs.~\ref{ac}\,c) and d) show. If the momentum $Q$ is large enough the extremum appears already above the critical temperature as seen in Fig.~\ref{acinv}. The condensate part vanishes at $T_{\rm C}$ and the gas part is the only contribution above
. For large total momentum the Bose distribution is well approximated by the Boltzmann distribution. This allows us to calculate ${\cal G}_{\rm m}$ explicitly which yields that the extremum appears at $T_{\rm ex}\approx0.22\mbox{ }\hbar^2Q^2/4m$. In other words we have a mere thermal effect. When the mean motion characterized by the total momentum comes in resonance with the thermal motion we observe an extremum in the critical scattering length.

The ladder and quasi-particle approximation proposed in this paper cannot describe the dynamic formation or breaking of bound states. This is due to the fact that for the formation or destruction of a bound state an exchange of energy and momentum either with a third particle or with the medium has to be allowed. Therefore either more diagrams or a self-consistent spectral function would have to be included in the T-matrix.  This is the reason why the pole of (\ref{T}) lies on the real axis, i.e., has no imaginary part, and therefore the bound state has an infinite life time. On the other hand Figs.~\ref{ob}\,a), b) and e) show that in the normal state $|\omega_{\rm B}|$ is of the order of the thermal energy, for $a_0<-a_{\rm C}$ and for $a_0\gg a_{\rm C}$, i.e., near the resonance. This makes the bound states very instable towards collisions with medium particles and therefore limits their life time. However experimental experience shows, that bound-state and cluster formation has even to be suppressed by decreasing the density to be able to directly investigate Bose condensation \cite{K02}. If Bose condensation shall be reached by decreasing the temperature, then bound states are always possible before $T_{\rm C}$ is reached for any interaction, as Fig.~\ref{ac}\,c) shows. The only exception is the ideal gas case, i.e., $a_0=0$.

The binding energies at typical conditions for some elements used for Bose-condensation experiments are compiled in table~\ref{tab}. These data show that the scattering length can be positive and its absolute value can be of the magnitude or even less than $a_{\rm C}$. In these cases the binding energy is more than two orders of magnitude higher than the thermal energy, i.e., the bound state is stable, although $|\omega_{\rm B}|$ is too small compared with experimental values of bound states near the continuum edge. For example the energy of the last vibrational state below the continuum edge for sodium is \mbox{$(-13100\pm900){\rm neV}$ \cite{STLEKT00}}. The reason for that difference is that the contact interaction is a low energy approximation. One would expect an increase of the binding energy for an increasing potential range. Table~\ref{tab} shows further that for stable bound states, i.e., $|\omega_{\rm B}|\gg T$, the influence of the medium on the binding energy is negligible. On the other hand, since a finite-range potential would stabilize the bound states a medium influence on them may be measurable.

\begin{figure}[t]
\psfig{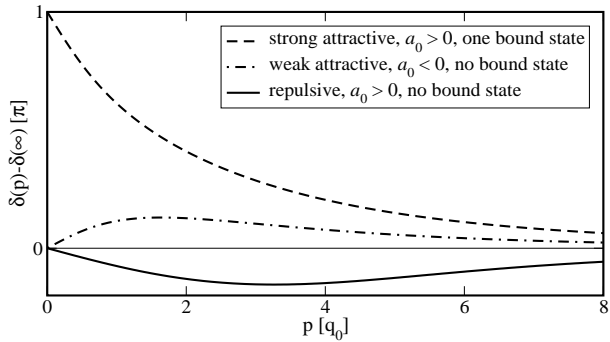} 
\caption{Scattering phase versus momentum for a finite-range two-particle interaction with one possible bound state}
\label{phase}
\end{figure}

To summarize, it has been shown that the contact interaction produces always attractive forces, if one postulates a finite scattering length. A bound state appears as soon as the interaction is strong enough. Due to the Bose enhancement, many-body effects support the formation of bound states while thermal fluctuations and the motion relative to the medium hinder this formation. The model describes the experimental experience that bound states and cluster formation appear before the Bose condensation. The calculations show that for bosons at finite density bound states are possible on both sides of a Feshbach resonance but they are quite instable near the resonance, too. To find better agreement with the experiment and to make predictions for future experiments, the model has to be improved to describe an interaction with finite range and bound states with finite life time. Our T-matrix approximation shows that, in contrast to the pseudopotential ansatz, the scattering length is not necessarily proportional to the interaction strength and that whether the interaction is attractive or repulsive does not follow inevitably from the sign of the scattering length. Whether an interaction is repulsive or attractive can be found from the sign of the scattering phase, the slope of which for small momenta is related to the scattering length, as shown in Fig.~\ref{phase}. The contact interaction can only describe the strong and weak attractive cases while the pseudopotential can only describe the weak attractive and repulsive cases. Therefore the sign of the scattering length alone is not decisive whether the interaction is attractive or repulsive. Otherwise one would have to face the paradoxical situation that bound states appear also for a repulsive interaction. Considering the BEC-BCS transition it seems therefore that one has to include inevitably the effect of a finite potential range in order to be able to describe bound states, pairing and BEC correctly.

The discussions with R.~Zimmermann are gratefully acknowledged. This work was supported by the German DAAD and Czech research plan MSM 0021620834, by DFG Priority Program 1157 via GE1202/06 and the BMBF and by European ESF program NES.

\bibliography{phasebind}

\end{document}